\documentclass[a4paper,prc,unsortedaddress,superscriptaddress,twocolumn,floatfix]{revtex4}
\usepackage{graphicx}		
\usepackage{bm}			
\usepackage{amsmath}
\usepackage{url}
\usepackage{xcolor}
\usepackage[normalem]{ulem}
\newcommand{\nscl}{National Superconducting Cyclotron Laboratory,
Michigan State University, East Lansing MI, USA}
\newcommand{\msuChem}{Department of Chemistry,
Michigan State University, East Lansing MI, USA}
\newcommand{\msuPhys}{Department of Physics and Astronomy,
Michigan State University, East Lansing MI, USA}
\newcommand{\witPhys}{Department of Physics,
Wittenberg University, Springfield OH, USA}
\begin{document}

\title{Measurement of the $^{20}$F half-life}

\affiliation{\msuPhys}
\affiliation{\nscl}
\affiliation{\witPhys}
\affiliation{\msuChem}

\author{M.~Hughes}
\affiliation{\msuPhys}\affiliation{\nscl}
\author{E.A.~George}
\affiliation{\witPhys}
\author{O.~Naviliat-Cuncic}
\altaffiliation[Corresponding author: ]{naviliat@nscl.msu.edu}
\affiliation{\msuPhys}\affiliation{\nscl}
\author{P.A.~Voytas}
\affiliation{\witPhys}
\author{S.~Chandavar}
\affiliation{\nscl}
\author{A.~Gade}
\affiliation{\msuPhys}\affiliation{\nscl}
\author{X.~Huyan}
\affiliation{\msuPhys}\affiliation{\nscl}
\author{S.N.~Liddick}
\affiliation{\nscl}\affiliation{\msuChem}
\author{K.~Minamisono}
\affiliation{\msuPhys}\affiliation{\nscl}
\author{S.V.~Paulauskas}
\affiliation{\nscl}
\author{D.~Weisshaar}
\affiliation{\nscl}

\date{\today}
           
\begin{abstract}
The half-life of the $^{20}$F ground state
has been measured using a radioactive beam implanted
in a plastic scintillator and recording $\beta\gamma$
coincidences together with four CsI(Na) detectors.
The result, $T_{1/2} = 11.0011(69)_{\rm stat}(30)_{\rm sys}$~s,
is at variance
by 17 combined standard deviations with the two most
precise results. The present value revives the poor consistency of
results for this half-life
and calls for a new measurement, with a technique having different
sources of systematic effects, to clarify the discrepancy. 
\end{abstract}

\maketitle
\section{Introduction}
\label{sec:intro}

The particular decay properties of the $A=20$ isospin triplet have
attracted considerable attention 
as being advantageous for performing
correlation measurements in $\beta$ decay that test the strong
form of the principle
of conservation of the vector current \cite{Gre85}. The comparison
of $ft$ values from $^{20}$F and
$^{20}$Na has been used for tests of mirror symmetry and
searches for second-class currents \cite{Wil70}.
In nuclear astrophysics, the lifetime of $^{20}$F plays a role
in the evolution of stars
with masses in the range 8-12$M_\odot$ which become electron-capture
supernovae \cite{Nom13}. The collapse is triggered by the loss of
electron pressure support
via the sequence $^{20}$Ne$(e^-,\nu)$$^{20}$F$(e^-,\nu)$$^{20}$O
on the very abundant $^{20}$Ne nuclear species \cite{Lan14}.
The precision requirements on the $^{20}$F lifetime for these
two domains
are vastly different. Whereas for astrophysical calculations of
electron capture rates a relative uncertainty of 10\% would be
sufficient \cite{Mar14}, mirror symmetry tests will ultimately be
limited by
the accuracy in the determination of the statistical rate function, $f$.
For the most favorable decays, this can be determined with an accuracy
of few $10^{-4}$ \cite{Hay18}.

\begin{figure}[!htb]
\centerline{
\includegraphics[width=\linewidth]{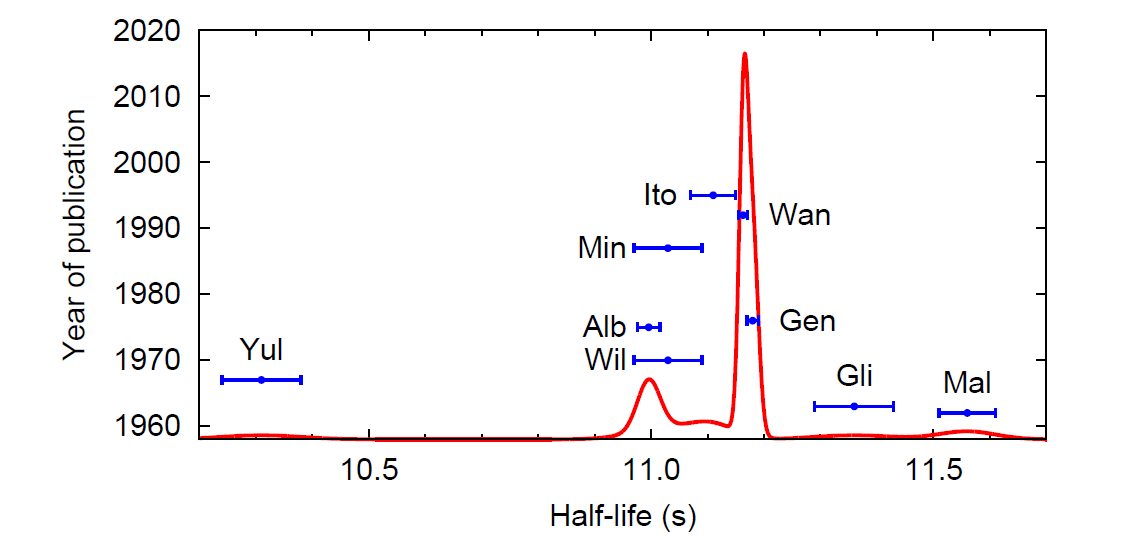}}
\caption{(Color online) Measurements of the $^{20}$F half-life prior
to the present work along with
their year of publication. The red curve is an ideogram obtained 
from a sum of Gaussians centered at the mean values and having a weight
$(1/\sigma_i^2)$, where $\sigma_i$ is the uncertainty of the
measurement. The labels correspond to: Mal~\cite{Mal62}, Gli~\cite{Gli63},
Yul~\cite{Yul67}, Wil~\cite{Wil70}, Alb~\cite{Alb75}, Gen~\cite{Gen76},
Min \cite{Min87}, Wan~\cite{Wan92} and Ito~\cite{Ito95}.}
\label{fig:ideogramBefore}
\end{figure}

The $\beta$ decay of $^{20}$F ($E_{\rm max} = 5.4$~MeV) occurs almost
exclusively (99.99\%) to
the first excited state in $^{20}$Ne which subsequently decays with
the emission of a 1.63~MeV $\gamma$-ray.
The $^{20}$F lifetime has therefore been measured by either detecting 
$\beta$ particles in singles, $\gamma$ rays in singles, or both in
$\beta\gamma$ coincidences.
The adopted value for the $^{20}$F half-life,
$T_{1/2} = 11.163(8)$~s \cite{Til98}, arises from a single
measurement \cite{Wan92} which detected $\gamma$ rays in singles
with a Ge(Li) detector. The value is consistent
with a previous result \cite{Gen76}  obtained by counting $\beta$
particles in singles using a magnetic spectrometer.
However, the adopted value \cite{Til98} does not reflect the spread
among the values previously measured. Figure~\ref{fig:ideogramBefore}
shows results from measurements of
the half-life having a total uncertainty, $\sigma_i$, which is at most 10
times larger than the most precise result of Ref.~\cite{Wan92}.
The red curve is a sum of Gaussians, weighted by $(1/\sigma_i^2)$,
and is dominated by the two most precise results. There appears to be
no correlation between the reported values and the experimental
techniques, whether these used $\beta$ particles in
singles \cite{Alb75,Gen76,Min87},
$\gamma$ rays in singles \cite{Gli63,Yul67,Wan92,Ito95}, or $\beta\gamma$
coincidences \cite{Mal62}. The short description in Ref.~\cite{Wil70}
suggests that the measurements were carried out with $\beta$
particles and $\gamma$ rays in singles, over different time intervals. 
The poor statistical consistency of the results can be quantified
by a fit of the nine values with a constant, which
gives $T_{1/2} = 11.1521(58)$~s with $\chi^2/\nu = 37.5$. 

The present work reports a high statistics measurement of the $^{20}$F
half-life performed by counting $\beta\gamma$ coincidences. The experiment
was carried
out in the framework of $\beta$-decay studies of $^{20}$F. Although
the settings of the main experiment, in particular the duration of the
decay time windows, were not optimized for the half-life measurement,
the conditions were particularly clean in terms of background,
and a number of ancillary diagnostics
were available to control possible systematic effects. 
A preliminary progress report of the work reported here has been 
presented elsewhere \cite{Hug17}.

\section{Experimental conditions}
\label{sec:setup}

The experiment was performed at the National Superconducting
Cyclotron Laboratory (NSCL) at Michigan State University. 
A primary $^{22}$Ne beam was accelerated to 150 MeV per nucleon
by the Coupled Cyclotron Facility and impinged on a 188~mg/cm$^2$
thick Be target where the $^{20}$F was produced by projectile
fragmentation.
The secondary beam was analyzed by the A1900 fragment
separator \cite{Mor03} and was directed to the experimental area.
During beam tuning, the beam purity was measured to be $99.4\%$
with a $4$~mm wide aperture at the focal plane of the A1900
separator. The only observed radioactive
contaminant was $^{19}$O, at a level of 0.23\%.
During regular measurement runs, the aperture at the
focal plane of the fragment separator was reduced to $1$~mm ($\pm 0.5$~mm),
and the amount of the $^{19}$O contaminant was reduced to the level
of 0.06\%.

\begin{figure}[!hbt]
\centerline{\includegraphics[width=\linewidth]{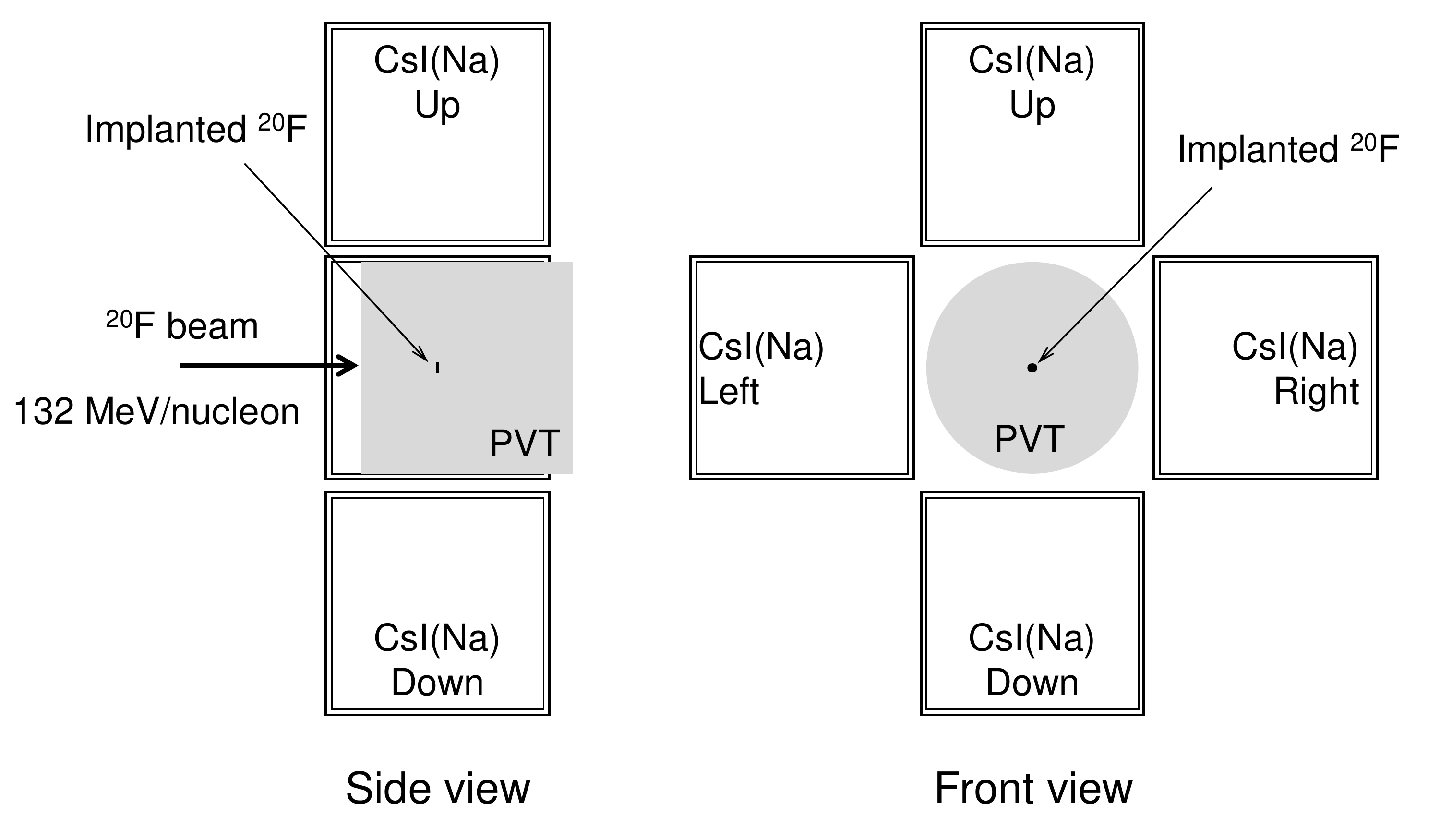}}
\caption{Layout of the experimental setup showing a side view (left) and
front view (right) of the detector arrangement around the position
where the beam stops. The PVT detector is offset downstream so that
the position of the implanted $^{20}$F is centered relative to the CsI(Na)
detectors.}
\label{fig:setup}
\end{figure}

In the experimental area, the $^{20}$F ions exited the beam pipe
through a 75~$\mu$m thick Zr vacuum window
and were implanted into a \O$7.6\times 7.6$~cm$^2$ EJ-200
polyvinyltoluene (PVT) plastic scintillator detector
(Fig.~\ref{fig:setup}). The beam energy before the PVT detector
was 132~MeV/nucleon.
Beam transport calculations using the LISE++ code \cite{lisePP}
indicated that
the beam was implanted at a mean depth of 3.02~cm inside the PVT and
produced a range straggling $\pm 0.6$~mm wide around the mean depth position.
The PVT was surrounded by four
$7.6\times 7.6\times 7.6$~cm$^3$ CsI(Na) modules from the CAESAR
array \cite{Wei10} for the detection of the $\gamma$ rays.
The transverse beam dimensions were measured with a position sensitive
parallel plate avalanche
counter (PPAC), placed in vacuum 40~cm upstream from the implantation
detector and removed during regular runs. 
The beam shape observed by the PPAC was elliptical,
8~mm wide and 6~mm high at full width tenth of maximum (FWTM).
Between the PPAC and the implantation detector the beam was convergent.
From the beam transport magnification and the measurements in the PPAC,
the actual size of the beam spot at the stopping location was deduced
to be 3.6~mm in the horizontal and 3.4~mm in the vertical directions
(FWTM).
The range of 5.4~MeV electrons in PVT is 2.75~cm. This means that $\beta$
particles from $^{20}$F decay cannot escape from the detector in
any direction.

A pulse generator was used to monitor the gain stability of the
PVT detector. The generator drove an external light emitting diode
(LED) linked to two parallel outputs via optical fibers.
 One of the outputs was connected to a Plexiglas ring which
served to couple the PVT detector to its photomultiplier tube (PMT)
and the other one was directed to an external Si PIN photodiode. 
The generator was operated at a trigger rate of 500~Hz
and produced two pulses of slightly different amplitude per
trigger, both well above the $\beta$-spectrum end-point,
and separated by 136~$\mu$s.

The CsI(Na) and PVT detectors were calibrated using, respectively, the
$\gamma$ lines and the Compton edges from
$^{22}$Na, $^{60}$Co and $^{137}$Cs sources, and the responses 
were found to be linear over the energy range covered by the sources.
The energy resolution of the four CsI(Na) detectors at 1.63~MeV was
between 7.3 to 7.9\% FWHM.
In the absence of beam, the ambient background rate of the CsI(Na)
detectors was about 250 counts per second (cps) and was less than 30 cps
for the PVT detector.

\section{Data acquisition and measurement sequence}
\label{sec:daq}

The data acquisition was based on the implementation of the
digital NSCL system \cite{Pro14} using three 250 mega-samples
per second Pixie-16 digitizing modules from the XIA company. 
The signals from the five detectors, i.e. the PVT and the four CsI(Na),
were sent to the first module.
The second module received the five signals from the PPAC, and
one signal from a Si detector used during beam tuning. 
Both Pixie modules also received signals associated with
the beam-on start, the beam-off start, an additional 100 Hz pulser,
and the signal from the Si PIN photodiode for monitoring. 
For each input channel in these two modules, the digitizer provided
the time stamp and an energy conversion using a trapezoidal
filter \cite{Pro14}.
The third Pixie module was used to digitize the waveforms from the PVT
detector over a 400~ns wide window. The waveforms
have been used to check the energy filter of the digitizer but have
otherwise not been specifically exploited for the half-life measurement
reported here.
The clock source generating the time stamping and signal
sampling in the Pixie-16 modules uses an EPSON SGR-8002JC-PCB
programmable crystal oscillator which has a frequency stability
of $\pm 5\times10^{-5}$ within a temperature range from
$-20$ to $70^\circ$C.

The Pixie modules continuously digitize the incoming waveforms but
do not record information until a threshold is crossed. Once the
threshold is crossed, the FPGAs in the Pixie system report an estimate of
the event energy based on particular sums of the samples of the waveform
at the input, following a trapezoidal filtering
of the signal. From the time the threshold is crossed until the
estimate is finished, no further triggers are acknowledged. However,
if the input waveform is due to the close overlap of more than one
signal, the presence of a second signal will affect the value of the
resulting energy estimate. This is essentially a pile-up event and
the value of the resulting energy estimate depends on the parameters
of the trapezoidal filter and on the relative timing between the
signals. The two signals are reported as one with an energy
estimate somewhere between that of the first signal alone and the
sum of the two signals, depending on the time difference between the
signals. The contribution of such pile-up events depend on the
particular energy cuts used and on the shape of the energy spectrum.
In the analysis described below, events lost from the summing region
were dealt with through the dead time correction and those added to
the summing region were dealt with through a pile-up correction.
There is no other known event loss in
this system at the rates of the experiment.

The time structure of a cycle consisted of a ``beam-on'' interval,
of 1.0, 1.1 or 1.67~s, during which the $^{20}$F beam was
implanted into the PVT detector, followed by a ``beam-off''
interval of 20, 30 or 60~s to measure the decay.
The beam chopping was performed by dephasing the radio-frequency
signal of one of the cyclotrons.
In order to reduce gain shifts in the PVT PMT resulting from
the large dynode currents during beam implantation,
the PMT HV was reduced (HV Inhibit) by a
factor of 2 during the implantation duration of the cycle.
This was applied for most of the runs (Table~\ref{tab:conditions}).
Other parameters such as the primary $^{22}$Ne$^{10+}$ beam intensity,
the durations of the
beam-on and beam-off windows, the high voltage on the PVT PMT
and the inhibit of the PMT high-voltage, were changed during the experiment
to check for possible systematic effects. The conditions are summarized
in Table~\ref{tab:conditions} and resulted in seven sets of runs listed
in chronologic order. Most of the runs in each set were one hour long. 

\begin{table*}[t]
\caption{Experimental conditions for the runs within the sets. For each
set, the table lists the
duration of the beam-on and beam-off
intervals, the high-voltage bias on the PVT, the setting of the
high-voltage inhibit, the primary beam intensity, the number of runs
and the total number of cycles.}
\begin{tabular}{lcclccrr}
\hline\hline
    &  Beam  &  Beam   &  PVT & HV  & Beam & & Total\\
Set &  on (s) & off (s) & HV (V) & inhibit  & intensity (nA) & Runs & cycles \\
\hline
1   & 1.67 & 30   & $-975$   & Off     & 30 & 9 & 915\\
2   & 1.67 & 30   & $-975$   & Off     & 93 & 2 & 77\\
3   & 1.67 & 30   & $-975$   & On      & 30 & 9 & 965\\
4   & 1.67 & 30   & $-975$   & On      & 93 & 11 & 1059\\
5   & 1.67 & 30   & $-856$   & On      & 93 & 10 & 1066\\
6   & 1.00 & 60   & $-800$   & On      & 93 & 1 & 63\\
7   & 1.10 & 20   & $-780$   & On      & 93 & 10 & 1604\\
\hline\hline
\end{tabular}
\label{tab:conditions}
\end{table*}

For the runs with a low primary beam intensity of 30~nA (electric),
the singles
counting rates 2~s after beam-off were about 3500~cps and 480~cps
for the PVT and CsI(Na) detectors, respectively. For runs
with the high beam intensity of 93~nA, these rates were typically 11500~cps
and 1200~cps.

\section{Sample spectra}
\label{sec:spectra}

\begin{figure}[!hbt]
\centerline{\includegraphics[width=\linewidth]{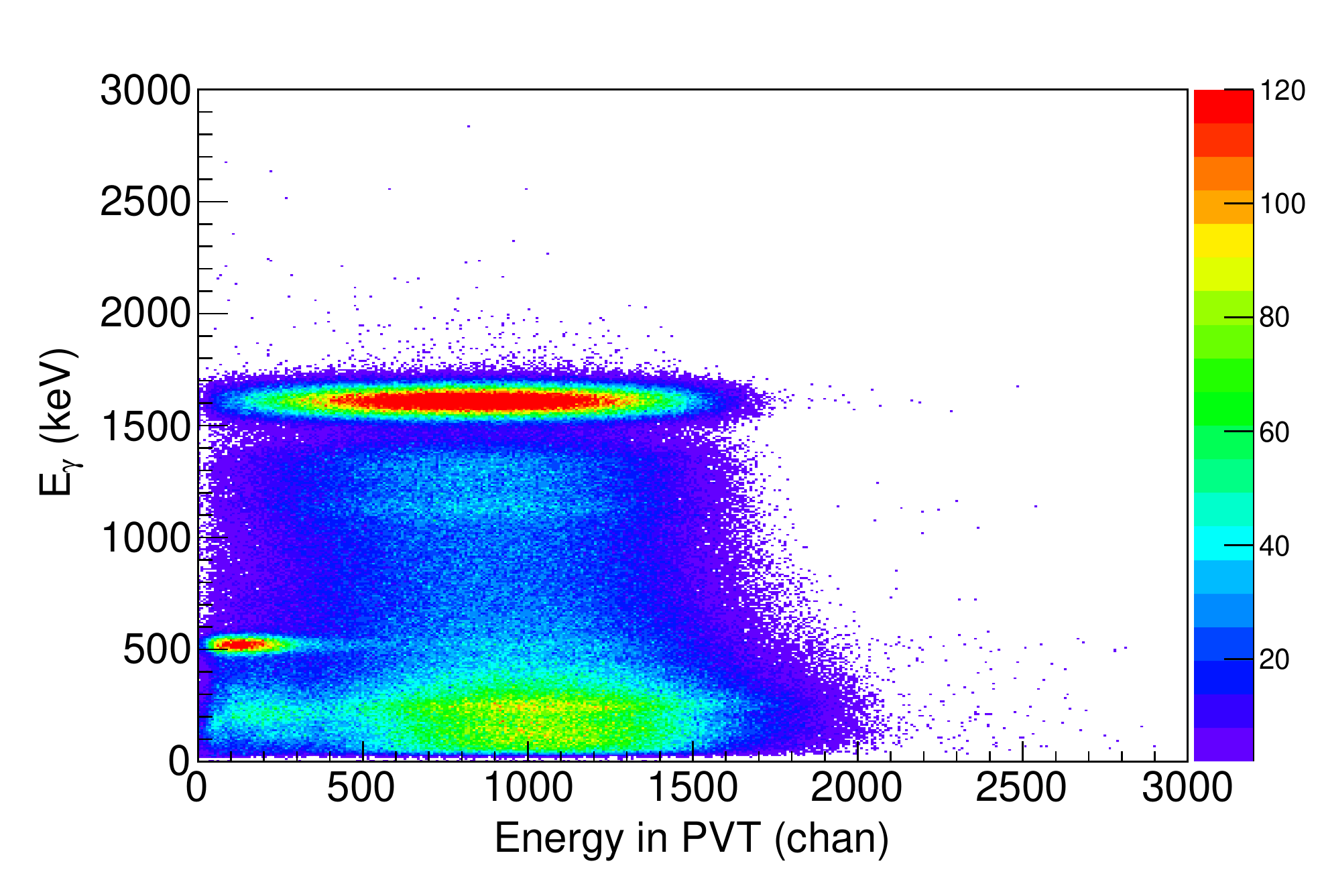}}
\caption{(Color online) Two-dimensional histogram of the energy in
the CsI(Na)-Right detector versus
the energy deposited in the PVT detector. See text for details.}
\label{fig:2D-plot}
\end{figure}

Figure~\ref{fig:2D-plot} shows a two-dimensional histogram
for a single run from set 7, of the energy deposited in the
CsI(Na)-Right detector  versus the energy in the PVT
detector, recorded during the beam-off interval.
The events were required to be within a 400~ns long software
event window generated by the first arriving signal.
The $\beta$ distribution
associated with the 1.63~MeV $\gamma$ coincidences is clearly visible.
There is no indication of transitions from contaminants giving rise
to $\gamma$ rays with energies larger than 1.63~MeV.
The analysis of events which are associated with the 0.511~MeV peak
indicated that those arise mainly from two $\beta^+$ emitters having
distinct half-lives and end-point energies. Their properties are
consistent with those of $^{10}$C and $^{11}$C decays, which can be
produced by reactions on $^{12}$C in the PVT. The production
of $^{10}$C
was confirmed by requiring a four-fold coincidence between the
implantation detector, two back-to-back CsI(Na) detectors, to record
pairs of 511 keV photons, and a third CsI(Na) detector to
identify the 718~keV $\gamma$ ray from $^{10}$B.
The identification of these distributions with $^{10}$C and $^{11}$C
was furthermore confirmed by a test in which the PVT detector was
replaced by a CsI(Na) detector and where no such distributions
were observed.

\begin{figure}[!htb]
\centerline{\includegraphics[width=\linewidth]{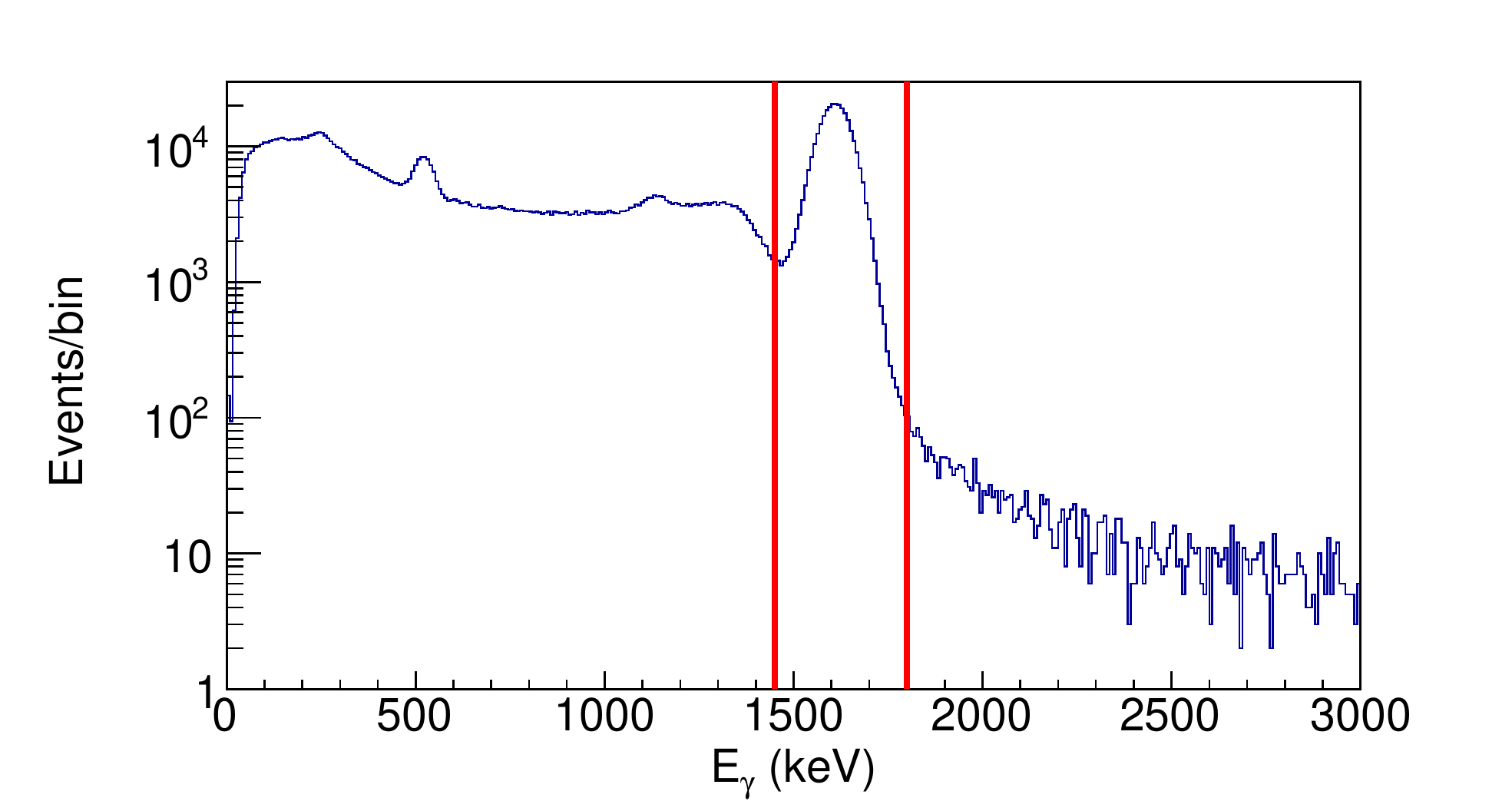}}
\caption{(Color online) Energy spectrum of $\gamma$ rays detected in
coincidence with a signal in the PVT implantation detector. The vertical
lines indicate the mean positions of the cuts applied for this detector.}
\label{fig:gamma_spec}
\end{figure}

\begin{figure}[!htb]
\centerline{\includegraphics[width=\linewidth]{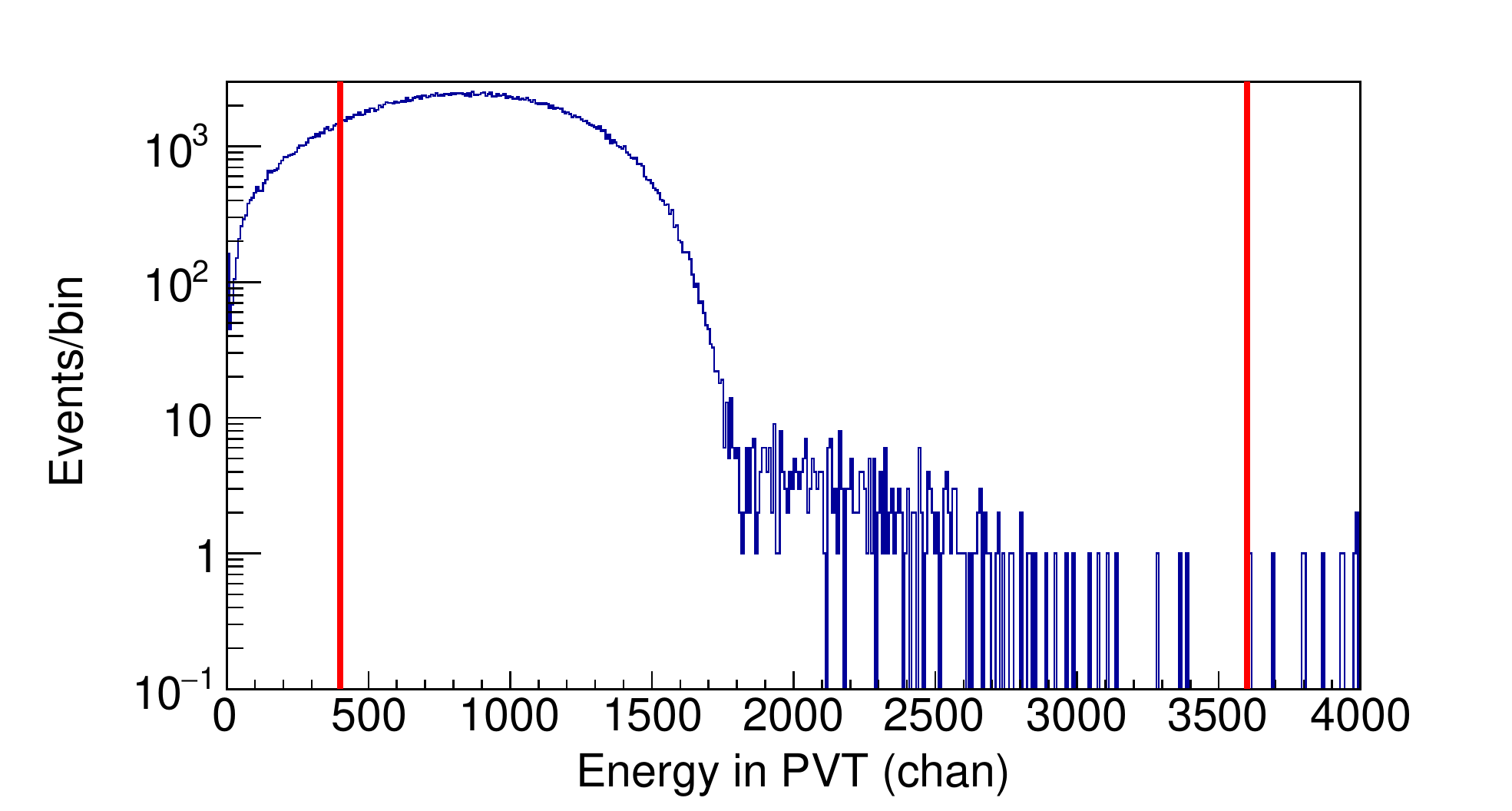}}
\caption{(Color online) Energy deposited in the PVT detector recorded in
coincidence with a signal in one of the four CsI(Na) detectors and
gated around the 1.63~MeV peak. The vertical lines indicate the
reference positions of cuts used for the final analysis in data set 6.}
\label{fig:beta_spec}
\end{figure}

Figure~\ref{fig:gamma_spec} shows the projection of Fig.~\ref{fig:2D-plot}
on the $\gamma$-energy (vertical) axis, without any condition on the
PVT energy. The vertical lines
show mean positions of cuts around the 1.63~MeV
peak which were varied in the analysis to test the stability of the
results. Towards lower energies from the
peak,
the spectrum shows the Compton edge and
the single escape peak. Monte-Carlo simulations indicated that the events
observed towards higher energies are dominated by bremsstrahlung produced by
electrons in the PVT detector. 
Figure~\ref{fig:beta_spec} shows a projection of
Fig.~\ref{fig:2D-plot} on the PVT energy, with the cuts shown in
Fig.~\ref{fig:gamma_spec}. The vertical lines in Fig.~\ref{fig:beta_spec}
indicate the position
of the cuts on the PVT energy that were also varied during the analysis
to test the stability of the results. 

\begin{figure}[!htb]
\centerline{\includegraphics[width=\linewidth]{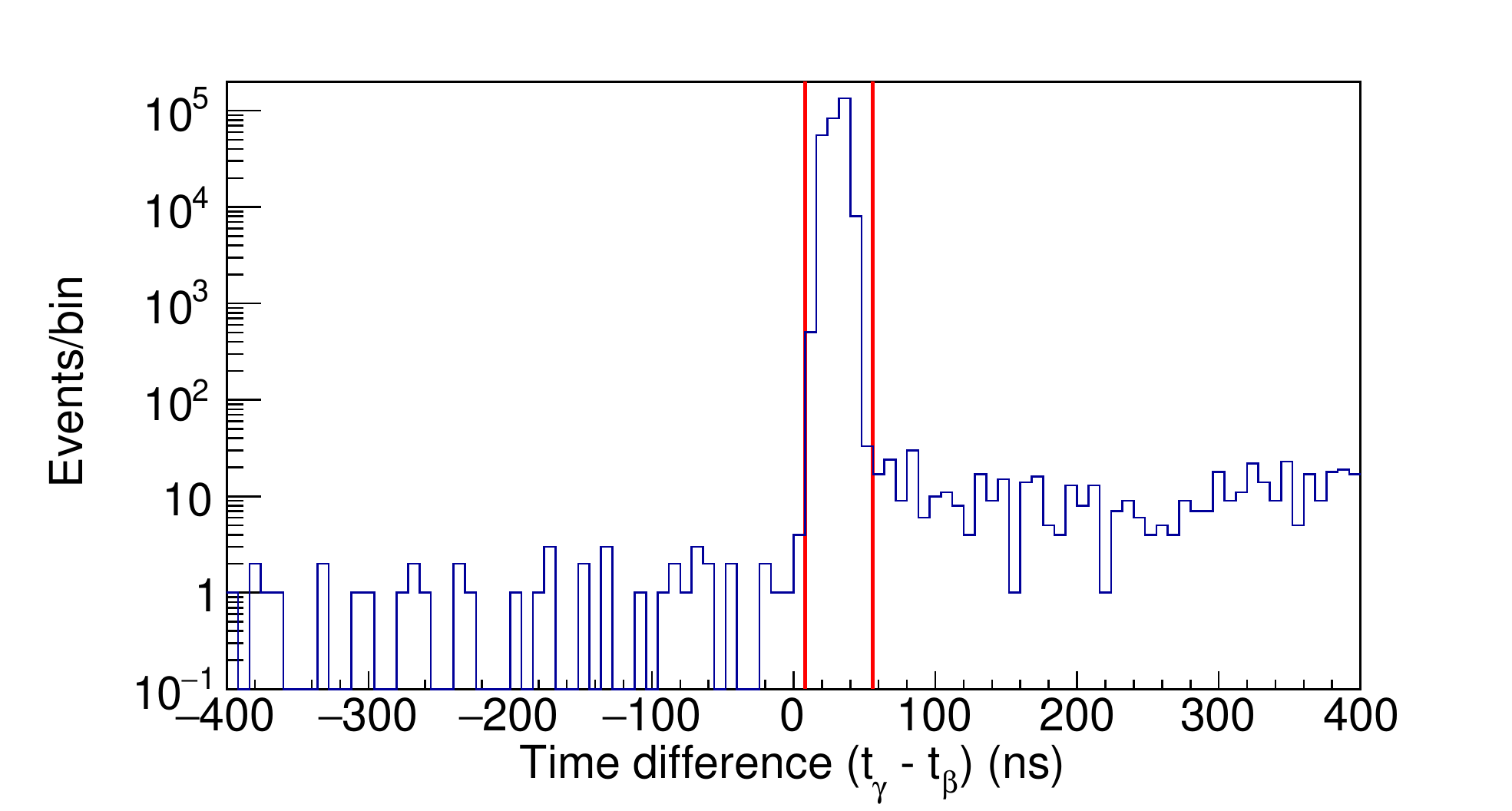}}
\caption{(Color online) Time difference between a signal from the PVT detector
and a signal from the CsI(Na)-Right detector obtained from events satisfying
the $\beta$ and $\gamma$ energy cuts.
The vertical lines indicate the narrowest cuts, $\pm 24$~ns around the
prompt peak, used in the analysis.} 
\label{fig:t_bg1}
\end{figure}

Decay histograms were built from
events in the 1.63~MeV peak window detected in coincidence with the
PVT detector.
Each CsI(Na) detector produces a statistically independent decay spectrum
resulting in 208 spectra from 52 runs distributed among the seven sets
of Table~\ref{tab:conditions}.
Cuts were also applied to the time difference
between the PVT and the CsI(Na) signals such as to select events 
around the
prompt peak (Fig.~\ref{fig:t_bg1}). All events in Fig.~\ref{fig:t_bg1}
satisfy by construction the energy cuts on the $\beta$
and $\gamma$ spectra. Events located left from the 
peak correspond to accidental coincidences from ambient background.
The larger level of events located right from the prompt peak is produced
by $\beta\gamma$ coincidences between time uncorrelated signals
associated with two decays in the PVT detector occurring within the
dead time window of the PVT channel. For these events, the second
$\beta$ signal is
dead time suppressed by the first; the $\gamma$ ray of the second
decay falls within the energy window of a CsI(Na) module and is then
detected in coincidence with
the first $\beta$. The $\gamma$ ray associated with the first decay
is either detected as a Compton event by one of the other three CsI(Na)
detectors or goes undetected. More details are given in Sec.~\ref{subsec:pileup}.

Two types of energy cut methods were used on the energy spectra,
one with fixed positions for each set of runs listed
in Table~\ref{tab:conditions} and another where the cuts were
determined
individually for each run, relative to the position of the 1.63~MeV
peak centroid
and to the endpoint of the $\beta$ spectrum. This second approach 
accounted for gain drifts between runs of a given set such as to
have more similar conditions for each run. Such adaptive cuts do not
account, however, for rate dependent gain shifts that can occur during the decay
window. Both methods gave consistent results. 

With the cuts around the $\gamma$ peak (Fig.~\ref{fig:gamma_spec}),
the lower cut in the $\beta$ energy spectrum above the $^{10,11}$C
contaminants and the cuts around the peak in the time
difference spectrum (Fig.~\ref{fig:t_bg1}), the typical coincidence rates
between
the PVT detector and one CsI(Na) detector were 35~cps for the runs
with the reduced beam intensity and 120~cps for the runs with
higher beam intensity. The typical average ambient background coincidence
rate with the same cuts was $1.5\times10^{-3}$~cps.

\section{Data analysis}
\label{sec:analysis}

For each CsI(Na) detector, the decay spectrum was built
after imposing the cuts on the $\gamma$ energy spectrum,
on the $\beta$ energy spectrum,
and on the time difference between the $\beta$ and $\gamma$
signals. Events were binned every 0.25~s.
In order to avoid edge effects, the histogram
was fit from $1.5$~s after the beginning of the
beam-off window up to $1.5$~s before the end of the window.
The time used in the decay spectrum was obtained from
the time stamp of the PVT detector signal relative to the time stamp
of the beam-off signal.
The fitting function was of the form
\begin{equation}
f(t) = a\exp{(-t \ln{2}/T_{1/2})}
\label{eq:fit-function1}
\end{equation}
with $a$ and $T_{1/2}$ as free parameters. The effect of
accidental coincidences due to ambient background
is discussed in Sec.~\ref{subsec:background}.
The parameters were determined using the log-likelihood method.
Fits of comparable quality were obtained before and after the dead time
corrections of decay spectra.

\begin{figure}[!htb]
\centerline{\includegraphics[width=\linewidth]{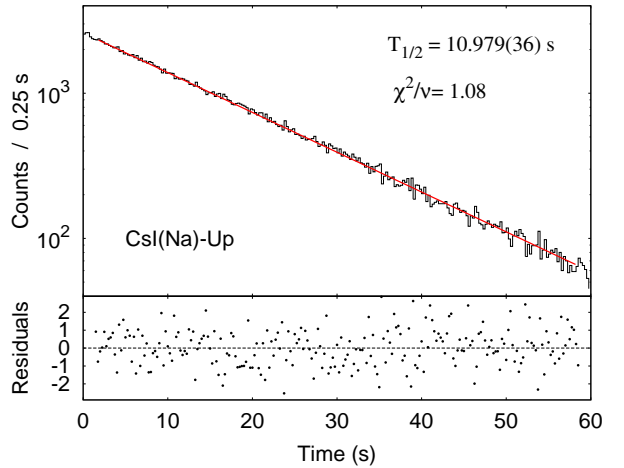}}
\caption{(Color online) Upper panel: Decay spectrum from the CsI(Na)-Up
detector (black histogram) and fitted function (red line).
Lower panel: Normalized residuals from the fit.}
\label{fig:decay_spec}
\end{figure}

The fit of a dead time corrected decay spectrum for the CsI(Na)-Up detector,
with a 60~s beam-off
window, is shown in Fig.~\ref{fig:decay_spec}. The associated
residuals are defined as $R_i = [f(t_i) - n_i]/\sqrt{n_i}$, with $n_i$ the
number of counts in bin $i$.
The other parameters for this particular run are given in
Table~\ref{tab:conditions}.

The values of half-lives resulting from fits having a p-value smaller
than 0.05 were excluded from the data sample. The fraction of such fits
was 0.06 and they are distributed over the full set
of runs and among the four detectors. This fraction
is consistent with expectations from pure statistical fluctuations.
To illustrate the weight of the data sets, the
results of the half-life for each set, averaged over all four detectors,
are presented in Table~\ref{tab:HLperSet}.
For reasons explained in Sec.~\ref{subsec:gainDrifts}, sets 1 and 2
were excluded from the data sample to calculate the final result.
From the remaining sets,
it is clear that sets 4 and 5 have the largest weight.

\begin{table}[!hbt]
\caption{Dead time corrected half-lives obtained for each data set from
the fits of decay spectra, averaged over the four CsI(Na) detectors.
For each run, the lower PVT energy cut was set just above the carbon
contaminants and the narrowest cuts were used
on the relative time between the $\beta$ and the $\gamma$ signals.}
\begin{tabular}{lcc}
\hline\hline
Set  &  $T_{1/2}$ (s)  &  $\chi^2/\nu$ \\
\hline
1   & 11.0303(175) & 0.88 \\
2   & 11.0100(242) & 1.06 \\
3   & 11.0258(152) & 0.94 \\
4   & 11.0033(~71) & 0.69 \\
5   & 10.9987(~77) & 1.12 \\
6   & 10.9735(253) & 0.05 \\
7   & 10.9869(108) & 0.74 \\
\hline\hline
\end{tabular}
\label{tab:HLperSet}
\end{table}

\begin{figure}[!htb]
\centerline{\includegraphics[width=\linewidth]{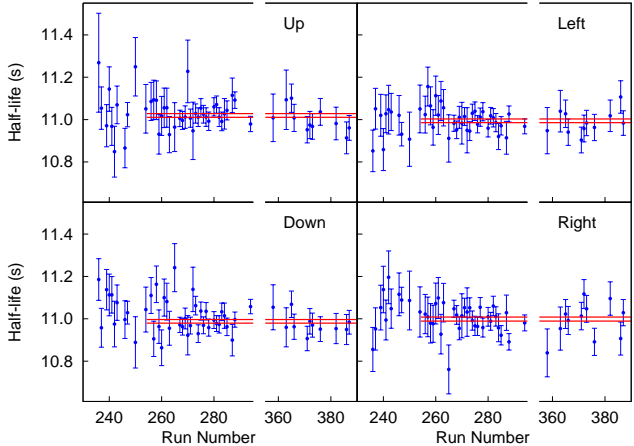}}
\caption{(Color online) Values of the half-lives obtained from fits of
the decay histograms for
the four CsI(Na) detectors as a function of the run number. The horizontal
bands indicate the $\pm 1\sigma$ limits of the fits
from data sets 3 to 7.}
\label{fig:halfLives}
\end{figure}

Figure~\ref{fig:halfLives} shows all individual values of half-lives
for the four detectors and for all runs after dead time correction.
The horizontal bands indicate for each detector the $\pm 1\sigma$
(statistical) limits of the fit of values from sets 3 to 7 having
a p-value larger than 0.05. The
values are summarized in Table~\ref{tab:HLperDetector} along with their
normalized $\chi^2$.

\begin{table}[!hbt]
\caption{Dead time corrected half-lives obtained from fits of values
from sets 3 to 7 for the four
CsI(Na) detectors with their corresponding normalized $\chi^2$.
For each run, the lower PVT energy cut was set just above the carbon
contaminants and the narrowest cuts were used
on the relative time between the $\beta$ and the $\gamma$ signals.}
\begin{tabular}{lcc}
\hline\hline
Detector  &  $T_{1/2}$ (s)  &  $\chi^2/\nu$ \\
\hline
Up   & 11.0195(88) & 0.62 \\
Left   & 10.9944(87) & 0.67 \\
Down   & 10.9880(85) & 1.02 \\
Right   & 10.9987(95) & 1.08 \\
\hline
Mean & 10.9999(44) & 2.44 \\
\hline\hline
\end{tabular}
\label{tab:HLperDetector}
\end{table}

A fit to the four values in Table~\ref{tab:HLperDetector}
gives the tabulated mean and
$\chi^2/\nu$, with a p-value~$=$~0.062.
A fit of all results from sets 3 to 7, without grouping them
first by detector, gives $T_{1/2} = 11.0001(44)$~s 
with $\chi^2/\nu = 0.87$ and a p-value~$=$~0.87.
If the fits with p-values smaller than 0.05 were included in
the sample, the half-life would change by $-3\times 10^{-4}$~s.
To account for the spread in the values obtained when grouping them
by detectors, which is due primarily to the Up detector,
the statistical uncertainty on the mean value was increased
by $\sqrt{\chi^2/\nu} = 1.56$ in the final result.

\section{Systematic effects}
\label{sec:systematics}

\subsection{Dead-time correction}
The shortest time differences between consecutive signals,
as measured from the time stamps, were $\tau_\beta = 464$~ns and
$\tau_\gamma = 656$~ns for respectively the PVT and CsI(Na) channels.
These determine the effective dead times of the
system for these channels. The time window of the software events,
of 400~ns, was chosen to be smaller than both of those effective
dead times. The measured coincidence rates, $r_{\beta\gamma}^m$, at a given
time after beam off were corrected according to
\begin{equation}
r_{\beta\gamma}^c =
\frac{1}{1 - r_\beta \tau_\beta} \cdot
\frac{1}{1 - r_\gamma \tau_\gamma}
r_{\beta\gamma}^m
\label{eq:deadtime}
\end{equation}
where $r_\beta$ is the singles rate in the PVT detector including events
from
the pulser and $r_\gamma$ is the singles rate in the associated coincident
CsI(Na) detector. In order to check for effects of intensity variations
during a run, the correction was applied to rates measured cycle by
cycle as well as to averaged rates measured over a run. Both methods gave
consistent
results since the data were taken under stable beam conditions. 

The relative size of the correction on the rates is
about $6\times10^{-3}$ at the
beginning of the decay histogram for the runs with the largest
beam intensity.
The comparison between the fitted values obtained
with and without dead time corrections showed that the
the central value of the half-life changed by
6 to 8~ms for the runs at low primary beam intensity
and by 31 to 36~ms for the runs at high primary beam intensity.
These corrections have been included in the results reported in
Tables \ref{tab:HLperSet} and \ref{tab:HLperDetector}.
Due to the 250~MHz sampling rate of the digitizers, there is an
uncertainty of $\pm 4$~ns on the measured dead times.
This induces a systematic
uncertainty of 0.24~ms on the half-life as listed
in Table~\ref{tab:errBudget}.

\begin{table}[!hbt]
\caption{Systematic effects considered in the error budget, with the
size of the effect on the half-life and the adopted uncertainty.
See text for details on the variations responsible for the
corrections and uncertainties.}
\label{tab:errBudget}
\begin{tabular}{lrr}
\hline\hline
Source &  Correction (ms)  &  Uncertainty (ms)  \\
\hline
Dead-time correction & 0.00\footnote{The mean value quoted in
Table~\ref{tab:HLperDetector} includes the dead-time correction.} & 0.24 \\
Oscillator stability & 0.00 & 0.80 \\
Uncorrelated events & 1.47 & 1.47 \\
Lower CsI(Na) cut & 0.00 & 0.15 \\
Upper CsI(Na) cut & 0.00 & 0.05 \\
Lower PVT cut & 0.00 & 2.32 \\
Binning & $-0.30$ & 0.30 \\
\hline
Systematic correction & 1.17 & 2.89 \\
\hline\hline
\end{tabular}
\end{table}

\subsection{Oscillator stability}

As indicated in Sec.~\ref{sec:daq}, the crystal oscillator has a
stability of $\pm 5\times 10^{-5}$. When applied to the duration of
the decay window, such a stability was observed to produce a variation
of $\pm 0.80$~ms on the extracted half-life. 

\subsection{Pile-up effects}
\label{subsec:pileup}
The dead time correction made through Eq.~(\ref{eq:deadtime}) accounts
for losses of coincidence events due to either $\beta$ or $\gamma$
suppressions by a previously occurring event within the dead time window
of the data acquisition channel. This corrects then for events removed
from the energy window defined by the cuts in Fig.~\ref{fig:gamma_spec}.
However, this energy window can also receive events
from the pile-up of two $\gamma$ rays having smaller energies or
from a $\gamma$ ray detected in coincidence with a
$\beta$ from a previously occurring event.

\begin{figure}[!htb]
\centerline{\includegraphics[width=\linewidth]{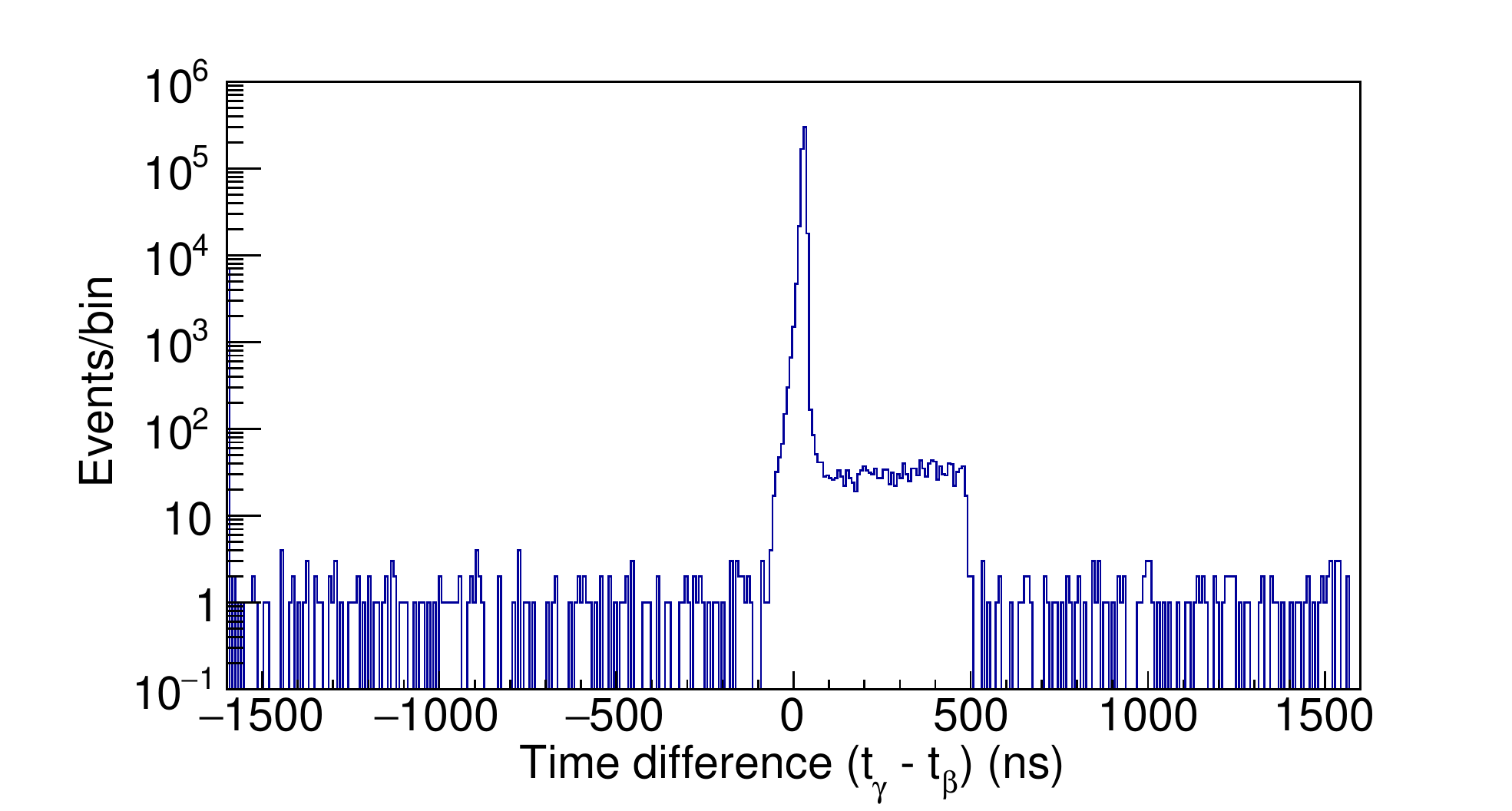}}
\caption{(Color online) Time difference between a signal from the
PVT detector
and a signal from the CsI(Na)-Left detector.
This is similar to Fig.~\ref{fig:t_bg1} but without the 400~ns
time condition between signals.}
\label{fig:t_bg2}
\end{figure}

As explained in Sec.~\ref{sec:analysis}, when
two $^{20}$F decay events occur during the
dead time of the $\beta$ channel they can give rise to the time
uncorrelated
events appearing at later times (right) from the prompt peak in
Fig.~\ref{fig:t_bg1}.
For clarity, Fig.~\ref{fig:t_bg2} displays a
spectrum of the relative time between the PVT and a CsI(Na) but
which has been
built without imposing the 400~ns wide software condition between
events.
The total duration of the plateau having a larger fraction of events
than the accidentals is $(\tau_{\beta} + \Delta t_{\beta\gamma})$
where $\Delta t_{\beta\gamma}$ is the mean time difference between
correlated $\beta$ and $\gamma$ signals for a given $\gamma$ channel.
The identification of events in the plateau
was checked by studying triple coincidence
events and comparing the results with expectations based on the detector
efficiencies. It was further tested by studying the decay time of
such events. Since the rate of uncorrelated events from two decays
occurring
during the dead-time window varies quadratically with the rate,
their decay time is expected to be half the decay time of
$^{20}$F.
This was confirmed by fitting the decay curve obtained
when setting a time window on the plateau of Fig.~\ref{fig:t_bg2}.
The contribution of such events has therefore an important impact
on the extracted half-life.
Because of the relatively low rate in the $\gamma$ detectors, the
pile-up with signals from ambient background, which are also
uncorrelated in time, has a much smaller contribution.

\begin{figure}[!htb]
\centerline{\includegraphics[width=\linewidth]{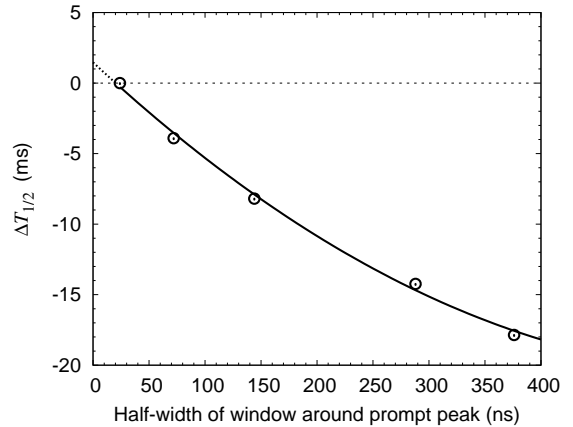}}
\caption{Systematic variation of the half-life obtained from sets
3 to 7 as
a function of half the width of the window used as cut around
the prompt peak in the time difference spectra (Fig.~\ref{fig:t_bg1}).
The value obtained for 24~ns is taken as
reference. The circles show the difference between the average
value resulting from the fits and the value obtained for 24~ns. The
solid black line is a quadratic description of the variation.
The extrapolation to zero width (dotted line) gives the size of
the correction.}
\label{fig:hl_vs_wtbg}
\end{figure}

In order to estimate the systematic effect due to the presence of such
uncorrelated events, the width of the time window used
on the relative time spectra (Fig.~\ref{fig:t_bg1}) has been increased
by more than a factor of 10, increasing thereby also the amount of
such events. It is clear from Figs.~\ref{fig:t_bg1} and \ref{fig:t_bg2}
that the number of
prompt events remains the same when increasing the window around the
peak.
The variation of the half-life as a function of the half-width of the
time window is shown in Fig.~\ref{fig:hl_vs_wtbg}. The narrowest window,
with $\pm 24$~ns cuts around the peak used in the analysis,
still contains a small fraction of such
uncorrelated coincidences. The extrapolation to zero of the variation
trend of the half-life gives
then the size of the correction, of $1.47$~ms.
Since this correction also depends on the lower cut on
the $\beta$ energy spectrum, its uncertainty is conservatively
taken as the value of the correction.

It is observed from Fig.~\ref{fig:hl_vs_wtbg} that the pile-up events
tend to decrease the half-life because, as explained above, their rate
varies quadratically with the decay rate of $^{20}$F.
A set by set analysis confirmed that the effect is
smaller for set 3 than for sets 4 to 7, as expected.
It is to note that the procedure described above also corrects for
uncorrelated events due to ambient background (Sec.~\ref{subsec:background})
which are
evenly distributed in the time coincidence spectrum of Fig.~\ref{fig:t_bg2}.
The combined systematic effect has therefore been labeled
``Uncorrelated events'' in Table~\ref{tab:errBudget}.

\subsection{Background}
\label{subsec:background}
The $\gamma$-energy cut (Fig.~\ref{fig:gamma_spec}) can
potentially allow the detection of background events. As mentioned above,
two other visible contributions in the 1.63~MeV peak window arise from
bremsstrahlung events, extending towards higher energies,
and from Compton events, due to the
detector resolution. Both types of $\beta\gamma$ events produce
prompt coincidences and they
have, furthermore, the same time signature as the main
$\beta\gamma$ events in the 1.63~MeV peak.

The lowest background level in the time decay spectrum
is determined by the contribution of accidental coincidences
from ambient background. It is delicate to determine such level
of accidentals in the decay spectrum by adding another
free parameter to the fit function in Eq.~(\ref{eq:fit-function1})
because of the correlations with the half-life and with the initial
activity. 
In order to reduce the correlations, an attempt was made
to simultaneously fit all four decay spectra
from a single run using a function of the form
\begin{equation}
g_i(t) = a_i \left[ \exp{(-t \ln{2}/T_{1/2})} + r_i \right]
\label{eq:fit-function2}
\end{equation}
where $a_i$ and $r_i$ are free parameters for each decay spectrum but
$T_{1/2}$ is a free parameter common to all spectra.
Such a constraint reduces the correlations between parameters for a
given detector but introduces correlations between the detectors.
The results from such analysis showed that the individual values of
the parameters $r_i$ are statistically consistent with zero and that
for the Left, Right and Down detectors there was a comparable amount
of negative and positive central values of these parameters.
For the Up detector, the fraction of
positive values was larger than of negative values, consistent with
the fact that the half-life for this detector is observed to be
larger when the background was assumed to be zero
(Table~\ref{tab:HLperDetector}), and consistent also
with the correlation with the other three detectors, which drive 
the value of the fitted half-life.

A similar analysis using Eq.(\ref{eq:fit-function2}) was performed
on sets of simulated data in order to study the sensitivity of
such a procedure to the background level and to confirm the
correlations and the statistical impact observed in the experimental
data.
The result showed that it is not possible to obtain a precise
determination of the background level over the relatively short
decay window, for each detector and for each run, when such level
is too small.

Alternatively, the level of accidental background can
be independently measured
for each detector by considering the events located left from
the prompt peak in Fig.~\ref{fig:t_bg2}. The analysis of all
decay spectra was then performed by determining first the background
level from those events in the time difference spectrum
and then fixing this level in the fitted function.
This analysis showed that the systematic error made by assuming
no background in the fitting function of Eq.(\ref{eq:fit-function1})
was 0.2~ms when using the $\pm 24$~ns time cuts on the time
difference spectrum.
The analysis also showed that, on average, for the same time window
on the relative time spectrum, the amplitude of the accidental
background relative
to the initial activity in the decay spectrum is $7.6\times 10^{-6}$.
This is one of the major advantages of the measurement described
here. Although the measurement is performed over a relatively short
time window, the level of accidental background is very low
due to the coincidence condition.

The studies described above are useful to understand the effect of
the ambient background separated from the pile-up.
As discussed in Sec.~\ref{subsec:pileup}, the correction due to the
time uncorrelated events shown in Fig.~\ref{fig:hl_vs_wtbg} includes
both, the pile-up events which give rise to the plateau
right from the prompt peak as well as the accidental background events
which are uniformly distributed.  Their effects on the half-life have
opposite signs but the pile-up events produce an effect which is a
factor of 8 larger, consistent with their distribution in
Fig.~\ref{fig:t_bg2}.
It is therefore not necessary to add an independent correction and
an associated uncertainty due to the level of accidental background.

\subsection{Gain drifts}
\label{subsec:gainDrifts}

\begin{figure}[!htb]
\centerline{\includegraphics[width=\linewidth]{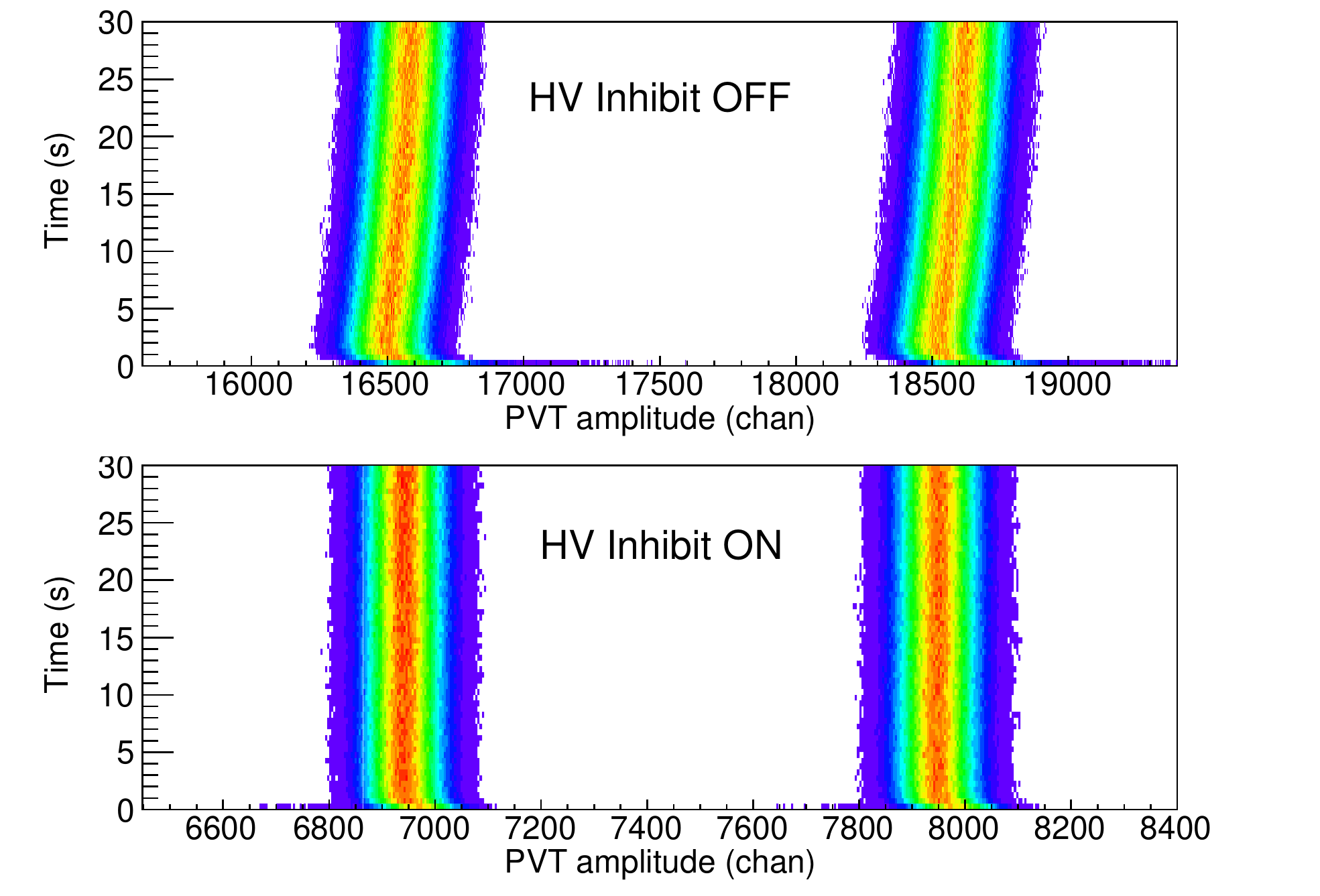}}
\caption{Variation of the amplitude of the two pulser signals
during the decay time. Upper panel: variation observed in a run from set
1 for which the HV inhibit was OFF.
Lower panel: variation from a run from set 5 with the HV inhibit ON.
The horizontal scales have been adjusted so that the relative
positions and widths
of the distributions are similar.} 
\label{fig:pulser_vs_time}
\end{figure}

The amplitudes of the PVT signals induced by the LED
were inspected during the decay window. For sets 1 and 2 of
Table~\ref{tab:conditions}, for which the HV inhibit was not active,
the LED-induced signals showed a relative gain drift of $4.8\times10^{-3}$
over the 30~s decay. This is reduced to $3.8\times10^{-4}$ in
the sets where the HV inhibit was active.
Figure \ref{fig:pulser_vs_time} shows the variation of the pulser
amplitudes for two runs, one with the inhibit OFF (upper panel)
and one with the inhibit ON (lower panel).

Because of the fixed lower cut on the PVT spectrum, the
presence of this time-correlated gain increase results in
a larger fraction of the energy spectrum being counted
at later times, producing a systematically larger half-life.

When using the lowest position of the lower cut in the PVT energy
spectrum (Fig.~\ref{fig:beta_spec}), the results obtained
from sets 1 and 2
do not display a significant difference in the values compared
to the other sets (Table \ref{tab:HLperSet}).
However, varying the lower cut in the PVT energy
has a significantly larger impact on the results from sets 1 and 2,
which cover run numbers up to 254 included, than for the runs in
the other sets. These systematic changes in sets 1 and 2 are
consistent with the measured gain drift.

In principle it is possible
to correct for this effect using the pulser
information. Such a procedure would anyway result in a larger 
uncertainty for these sets and we have opted to exclude
sets 1 and 2 from the data to extract the final result.

\subsection{Sensitivity to cuts}
\label{subsec:cuts}
 
The cuts on the CsI(Na) energy (Fig.~\ref{fig:gamma_spec}) were 
defined such as to reduce the sensitivity to possible gain drifts.
Off-line measurements have shown that, for the rate variations measured
by
the CsI(Na) detectors during the decay, the effect of rate-correlated
gain drifts is negligible. The lower and upper cuts on the CsI(Na)
detectors were independently varied by $\pm 5$~keV, corresponding
to a rate correlated gain variation which is 10 times larger than
expected for the actually measured rates.

For the PVT detector, the upper cut on the energy distribution was set
sufficiently high with respect to the main $\beta$ spectrum and the
pile-up events, and below
the position of the LED pulser signals (Fig. \ref{fig:beta_spec}).
The lower cut in the PVT energy is the most sensitive of all cuts
and, as mentioned above, was found to produce a systematic trend
for sets 1 and 2, consistent with the observed gain drift.

For lower cuts ranging from above the $^{10,11}$C contaminants up
to the
middle of the $\beta$ spectrum, there was no significant trend
observed in the values of the half-life for sets 3 to 7.
For the final analysis, the lower PVT energy cut was set at the
lowest end of this range.
The systematic uncertainty associated with the variation of the
lower PVT cut was taken as the maximal variations on the half-life
observed for sets 3 to 7, and was $\pm 2.32$~ms.

\begin{figure}[!htb]
\centerline{\includegraphics[width=\linewidth]{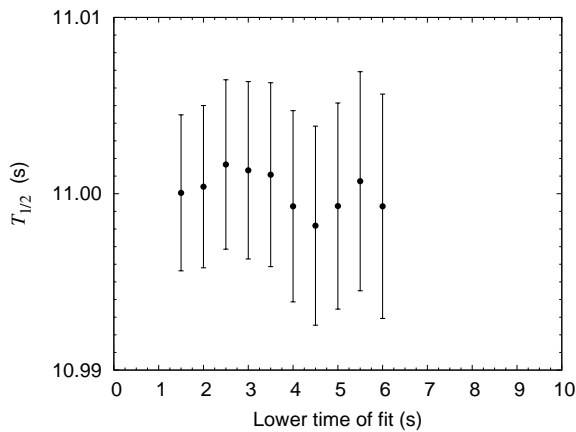}}
\caption{Mean value of the half-life obtained from the fits
of the decay spectra as a function of the lower time of the fit.
The error bars display only the statistical uncertainty.}
\label{fig:hl_vs_lowDecayCut}
\end{figure}

For the decay histograms, the lower 24 and
upper 24 bins, which span 6~s on each side, were independently
removed. No systematic trend was observed in the variation of the
central values. For illustration, Fig.~\ref{fig:hl_vs_lowDecayCut}
shows the values of the half-life as a function of the starting
time of the fit, indicating no systematic trend of the central
values with rate.

\subsection{Binning and fitting method}

The fits have been performed using log-likelihood estimators on
summed data.
Two independent analysis were performed using methods from two
packages: {\em Physica} \cite{Physica} and Root version 6.04.
When applied to the same data, the two methods gave identical
results.
The Root methods have been tested with simulated data,
having comparable statistics, fitting range and binning as those
used for the experimental data.
The Physica methods were also tested with experimental data,
by comparing the result from the fit of a summed histogram
with the result from the analytic solution of the maximum likelihood
estimator for the lifetime, calculated with unbinned events from the
data stream.
The analytic solution is given by the sample mean corrected by the
finite time of the measuring window.
No bias in the minimization methods at the current
level of precision was found.
The stability of the results has also been tested as a function of
the binning of the decay histogram. The final central value changed
by $-0.6$~ms when the number of bins was reduced by a factor of 2,
consistent with Monte-Carlo simulations. This
has been included in the list of systematic effects with a correction
of $-0.3$~ms and a systematic uncertainty of 0.3~ms.

\subsection{$^{20}$F Diffusion}

Any process reducing the number of nuclei in the sample with time,
other than $\beta$-decay, will result in a shorter half-life. We
have considered the possibility of F diffusion out of the PVT.
It is difficult to find Arrhenius coefficients which closely
correspond to actual experimental conditions, with F atoms
implanted in polyvinyltoluene at room temperature.
Based on fluorination work of fluorine gas in other polymer
films \cite{Shi75} it has been estimated that in 30~s, the
root mean square radial displacement is about 0.75~$\mu$m.
This estimate indicates that diffusion effects are negligible.

\section{Result and discussion}
\label{sec:result}

\begin{figure}[!htb]
\centerline{
\includegraphics[width=\linewidth]{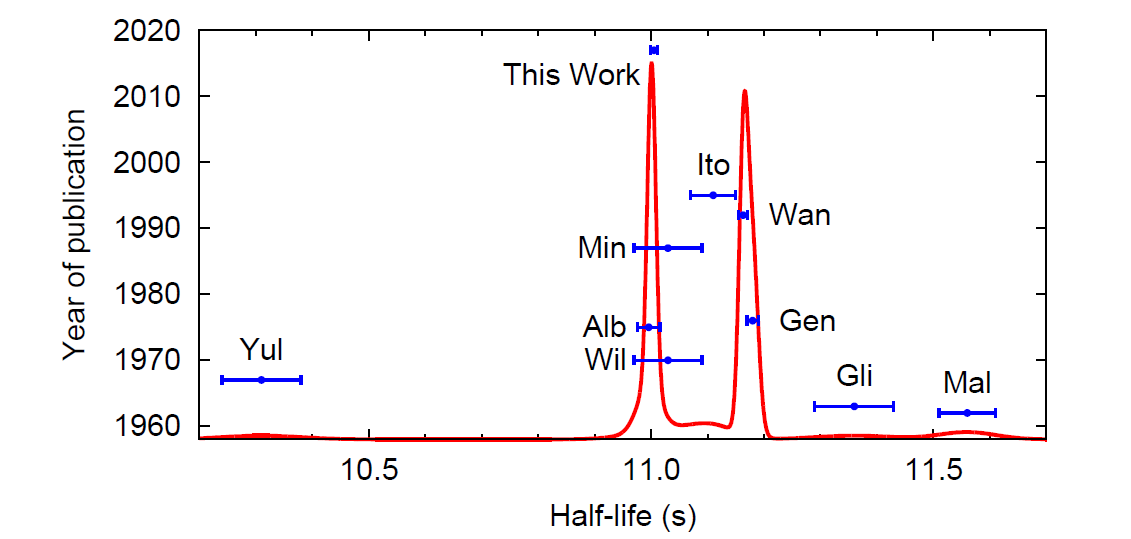}
}
\caption{(Color online) Measurements of the $^{20}$F half-life
including the result from the present work.
See caption of Fig.~\ref{fig:ideogramBefore} for details.}
\label{fig:ideogramAfter}
\end{figure}

With the scaling of the statistical uncertainty of the result from
Table~\ref{tab:HLperDetector} and the total systematic correction and error
listed
in Table~\ref{tab:errBudget}, the final result from this measurement is
\begin{equation}
T_{1/2} = 11.0011 (69)_{\rm stat} (30)_{\rm sys}~{\rm s,}
\label{eq:t12}
\end{equation}
and is listed along with previous results in Table~\ref{tab:listPastMeas}.
This value is consistent with those from
Wilkinson and Alburger \cite{Wil70} and from Minamisono \cite{Min87},
which are identical, and is also consistent with the value from
Alburger and Calaprice \cite{Alb75}. The weighted mean of the two most
precise results \cite{Gen76,Wan92}, which are mutually consistent and
dominate the current value,
gives $T_{1/2} = 11.1696(62)$~s. The value in Eq.(\ref{eq:t12})
is at variance by 17 combined standard deviations from
this weighted mean.
The impact of the value obtained from the present work on the set of
available results is shown in Fig.~\ref{fig:ideogramAfter} and is to
be compared with the prevailing situation
shown in Fig.~\ref{fig:ideogramBefore}. 

\begin{table}[!hbt]
\caption{Values of the $^{20}$F half-life obtained in previous
measurements along with the result from this work. The ``label'' in
the second column refers to those used in Figs.~\ref{fig:ideogramBefore}
and \ref{fig:ideogramAfter}.}
\begin{tabular}{lr}
\hline\hline
$T_{1/2}$ (s)  &  Label (year) Ref. \\
\hline
11.56(5) & Mal (1962) \cite{Mal62} \\
11.36(7) & Gli (1963) \cite{Gli63} \\
10.31(7) & Yul (1967) \cite{Yul67} \\
11.03(6) & Wil (1970) \cite{Wil70} \\
10.996(20) & Alb (1975) \cite{Alb75} \\
11.18(1) & Gen (1976) \cite{Gen76} \\
11.03(6) & Min (1987) \cite{Min87} \\
11.163(8) & Wan (1992) \cite{Wan92} \\
11.11(4) & Ito (1995) \cite{Ito95} \\
11.0011(75) & This work \\
\hline\hline
\end{tabular}
\label{tab:listPastMeas}
\end{table}

It is difficult to comment on eight among the nine previously
published results because they do not contain
detailed accounts of the measuring conditions and data analysis.
Only Wang {\em et al.} \cite{Wan92}
performed a dedicated measurement that attempted to resolve the then
existing
discrepancies. It is intriguing that the uncertainties quoted
there for the half-life (Table 1 of Ref.~\cite{Wan92}) are almost
identical when the background was either floating or fixed in the fits.
When adding a constant background as a free parameter, the correlation
between this parameter and the half-life increases the statistical
uncertainty on the half-life parameter. A simulation of decay
spectra with
similar conditions to those described in Ref.~\cite{Wan92} showed that
the uncertainty on the half-life increases by a factor of about 5
when the background parameter is left free as compared to when
it is fixed in the fits.
Even though the discussion lists a number
of systematic effects, the uncertainty quoted in Ref.~\cite{Wan92}
is purely statistical. 

\section{Conclusion}
\label{sec:conclusion}

Nine values of the $^{20}$F half-life have been reported
in the literature for which the uncertainty is at most 10 times
larger than the most precise result. For eight of them the information
about the experimental conditions and the data analysis is rather
scarce.
The measurement of Wang {\em et al.} \cite{Wan92} aimed at
resolving existing discrepancies among previously published
results.
The present work reported a new measurement of the half-life by
counting $\beta\gamma$ coincidences with a digital data acquisition
system which recorded the energies and the time stamps. Two major
advantages of the technique used here are: i) the dead-time of the
counting channels were
smaller than 660~ns and ii) the level of accidental coincidences due
to ambient background relative to the initial activity was
smaller than $8\times10^{-6}$. A detailed description of the
experimental conditions, of the data analysis, and of systematic effects
was given. The weight of the value obtained here
revives the poor consistency among existing results by adding tension
with the most precise results.
This calls for a new measurement of the half-life, with a technique
having different sources of systematic effects, in order to clarify the
discrepancy.

\medskip
\begin{acknowledgments}
We thank M.~Brodeur and X.~Fl\'echard
for fruitful discussions and T.~Chuna for his
assistance during the experiment.
This work was supported by the National Science Foundation under Grants
No. PHY-1102511, PHY-1506084, and PHY-1565546.
\end{acknowledgments}


\begin{thebibliography}{10}

\bibitem{Gre85}
L.~Grenacs,
Ann. Rev. Nucl. Part. Sci., {\bf 35}, 455 (1985).

\bibitem{Wil70}
D.H.~Wilkinson and D.E.~Alburger,
Phys. Rev. Lett., {\bf 24}, 1134 (1970).

\bibitem{Nom13}
K.~Nomoto, C.~Kobayashi, and N.~Tominaga,
Ann. Rev. Astron. Astrophys., {\bf 51}, 457 (2013).

\bibitem{Lan14}
K.~Langanke and G.~Mart\'{i}nez-Pinedo,
Nucl. Phys. A, {\bf 928}, 305 (2014).

\bibitem{Mar14}
G.~Mart\'{i}nez-Pinedo, Y.H.~Lam, K.~Langanke, R.G.T.~Zegers and C.~Sullivan
Phys. Rev. C, {\bf 89}, 045806 (2014).

\bibitem{Hay18}
L.~Hayen, N.~Severijns, K.~Bodek, D.~Rozpedzik, and X.~Mougeot,
Rev. Mod. Phys., {\bf 90}, 015008 (2018).

\bibitem{Til98}
D.R.~Tilley, C.M.~Cheves, J.H.~Kelley, S.~Raman, and H.R.~Weller,
Nucl. Phys. A, {\bf 636}, 249 (1998).

\bibitem{Wan92}
T.F.~Wang, R.N.~Boyd, G.J.~Mathews, M.L.~Roberts, K.E.~Sale, M.M.~Farrell, M.S.~Islam, and G.W.~Kolnicki,
Nucl. Phys. A, {\bf 536}, 159 (1992).

\bibitem{Gen76}
H.~Genz, A.~Richter, B.M.~Schmitz, and H.~Behrens,
Nucl. Phys. A, {\bf 267}, 13 (1976).

\bibitem{Alb75}
D.E.~ Alburger and F.P.~Calaprice,
Phys. Rev. C, {\bf 12}, 1690 (1975).

\bibitem{Min87}
T.~Minamisono,
Hyperfine Interact., {\bf 35}, 979 (1987).

\bibitem{Gli63}
S.S.~Glickstein and R.G.~Winter,
Phys. Rev., {\bf 129}, 1281 (1963).

\bibitem{Yul67}
H.P.~Yule,
Nucl. Phys. A, {\bf 94}, 442 (1967).

\bibitem{Ito95}
S.~Itoh, M.~Yasuda, H.~Yamamoto, T.~Iida, A.~Takahashi, and K.~Kawade,
in {\em {1994 Symposium on Nuclear Data}}, Tokai, Japan, March 1995.

\bibitem{Mal62}
S.~Malmskog and J.~Konijn,
Nucl. Phys., {\bf 38}, 196 (1962).

\bibitem{Hug17}
M.~Hughes, O.~Naviliat-Cuncic, P.~Voytas, E.~George, and X.~Huyan,
Bull. Am. Phys. Soc, {\bf 62}, S13.4 (2017).

\bibitem{Mor03}
D.J.~Morrissey, B.M.~Sherrill, M.~Steiner, A.~Stolz, and I.~Wiedenhoever,
Nucl. Instr. Meth. Phys. Res. B, {\bf 204}, 90 (2003).

\bibitem{Wei10}
D.~Weisshaar, A.~Gade, T.~Glasmacher, G.F. Grinyer, D.~Bazin, P.~Adrich,
  T.~Baugher, J.M.~Cook, C.A.~Diget, S.~McDaniel, A.~Ratkiewicz, K.P.~Siwek, and K.A.~Walsh,
Nucl. Instr. Meth. Phys. Res. A, {\bf 624}, 615 (2010).

\bibitem{lisePP}
O.B.~Tarasov and D.~Bazin,
Nucl. Instrum. Methods in Phys. Res. B, {\bf 376}, 185 (2016).

\bibitem{Pro14}
C.J.~Prokop, S.N.~Liddick, B.L.~Abromeit, A.T.~Chemey, N.R.~Larson, S.~Suchyta, and J.R.~Tompkins,
Nucl. Instr. Meth. Phys. Res. A, {\bf 741}, 163 (2014).

\bibitem{Physica}
Physica,
\url{http://computing.triumf.ca/legacy/physica/}.

\bibitem{Shi75}
J.~Shimada and M.~Hoshino.
J. Appl. Polym. Sci., {\bf 19}, 1439 (1975).

\end{thebibliography}

\end{document}